# Continuous Resilience in Cyber-Physical Systems of Systems: Extending Architectural Models through Adaptive Coordination and Learning

Elisabeth Vogel[1,2], Peter Langendörfer[1,2]

[1]Leibniz Institute for High Performance Microelectronics (IHP), Frankfurt (Oder), 15236, Germany
[2]Chair of Wireless Systems, BTU Cottbus-Senftenberg, Cottbus, 03046, Germany

*Abstract:*

*Cyber-physical systems of systems (CPSoS) are highly complex, dynamic environments in which technical, cybernetic and organisational subsystems interact closely with one another. Dynamic, continuously adaptable resilience is required to ensure their functionality under variable conditions. However, existing resilience architectures usually only deal with adaptation implicitly and thus remain predominantly static.*
*This paper addresses this gap by introducing a new Adaptive Coordination Layer (ACL) and conceptually redefining the Adaptation & Learning Layer (AL). The ACL acts as an operational control layer that detects risks in real time, prioritises countermeasures and coordinates them dynamically.*
*The AL is reinterpreted as a strategic-cooperative layer that evaluates the operational decisions of the ACL, learns from them, and derives long-term adjustments at the policy, governance, and architecture levels. Together, both layers operationalise the resilience principle of adaptation and combine short-term responsiveness with long-term learning and development capabilities.*
*The paper describes various implementation variants of both levels – from rule-based and KPI-driven approaches to AI-supported and meta-learning mechanisms – and shows how these can be combined depending on system complexity, data availability and degree of regulation.*
*The proposed architecture model no longer understands resilience as a static system property, but as a continuous, data-driven process of mutual coordination and systemic learning. This creates a methodological basis for the next generation of adaptive and resilient CPSoS.*

## 1. Introduction

Cyber-physical Systems of Systems (CPSoS) are highly complex, interdependent and dynamic environments in which numerous subsystems – technical, cyber and organisational – continuously interact with one another. Due to their size and heterogeneity, CPSoS are exposed to a multitude of risks, ranging from hardware and software errors to cyber-attacks and human error. In this context, simply identifying risks is not enough. Risk identification merely describes the current threat status of a CPSoS. However, such risk identification does not ensure that a CPSoS can withstand disturbances or adapt in such a way that the effects of future disturbances can be avoided or at least reduced.

Resilience engineering therefore plays a central role in ensuring that critical functions are maintained. It can be defined as an engineering discipline that focuses on enabling systems to anticipate, withstand, recover from and adapt to disturbances, ensuring that essential system functions remain available even under adverse conditions. Resilience engineering does not only address robustness and recovery but also emphasises continuous learning and adaptation across technical, organisational and operational layers.

Resilience goals such as anticipation, resistance, recovery, error analysis and adaptation form guiding principles for the design and operation of systems [1–4]. Countermeasures (CMs) are the operational levers that link risk identification with resilience goals: they reduce, control or compensate for the effects of disturbances. Without countermeasures, a CPSoS cannot be resilient despite good risk identification. Risks in CPSoS are often multi-dimensional and variable over time, so a structured approach to preventing or at least minimising the effects of such risks on a CPSoS is essential.

Two categories of countermeasures have dominated existing frameworks [5–11] to date:

- Universal measures that are effective system-wide and in the long term (e.g. redundancy, hardening, segmentation)
- Specific measures that target specific risks (e.g. firewalls against malware, fire protection systems in buildings)

However, these categories have a crucial limitation: adaptation is only dealt with implicitly. Learning processes, graceful extensibility and long-term adjustments are often implied within overarching concepts such as organisational culture or continuous improvement, without being considered as an explicit, measurable capability. As a result, existing resilience models remain predominantly static – focused on universal or specific countermeasures – while the ability to dynamically coordinate and further develop these measures is underrepresented.

Early initiatives in industry and the energy sector already show that adaptive coordination and learning mechanisms are beginning to enter real CPS and CPSoS applications. Pilot projects such as DEIS (Dependability Engineering Innovation for Industrial CPS) [5], CAPRI (Cognitive Automation Platform for Resilient Industry) [6], and the GRIP Grid Resilience and Innovation Partnership [7] demonstrate that dynamic adaptation is not only a conceptual requirement, but an emerging practical capability. These developments underline the need for architectural models that explicitly operationalise adaptation.

This paper addresses this gap by explicitly operationalising the resilience goal of 'adaptation'. To this end, we introduce a new Adaptive Coordination Layer (ACL) and redefine the existing Adaptation & Learning Layer (AL) to integrate adaptation more explicitly into resilience models:

1. **Adaptive Coordination Layer (ACL)** – an *operational, near-real-time* unit that continuously monitors risks and the effectiveness of countermeasures and dynamically activates, adapts or deactivates them.
2. **Adaptation & Learning Layer (AL)** – a *strategic, long-term* process level that integrates experience and insights, adapts policies and improves resilience capabilities over time. In contrast to previous literature, which predominantly treats adaptation as a downstream or cultural aspect, we define it here as an independent, structured and measurable level.

By formally distinguishing between Adaptive Coordination Layer and the Adaptation & Learning Layer, we bridge the gap between short-term operational coordination and long-term organisational learning. Together, these two layers provide a comprehensive framework for systematically embedding adaptation into CPSoS resilience architectures.

To this end, Section 2 describes the importance of resilience in CPSoS and provides a brief overview of existing resilience architectures. Section 3 discusses existing resilience engineering frameworks and shows how they implement the two groups of countermeasures (universal and specific CMs). Section 4 introduces the new Adaptive Coordination Layer and the redefinition of the Adaptation & Learning Layer. It also presents the integration of the proposed approach into existing resilience architectures. Section 5 presents concrete implementation proposals for the layers described in Section 4. Section 6 discusses the various configurations. Section 7 summarises the paper and provides an outlook on future work.

## 2. Background

CPSoS combine physical, cyber and organisational subsystems. Due to the close coupling and high dynamics, local disturbances can quickly spread to other parts and amplify system-wide. Resilience is therefore a key capability for ensuring that critical functions are maintained even under uncertain and changing conditions [1, 2, 8, 9].

In order to systematically capture and evaluate resilience in CPSoS, resilience engineering draws on a number of fundamental target areas. These resilience goals structure the capabilities a system needs to remain functional even in the face of disturbances and uncertainties.

### 2.1 Principles of Resilience Engineering - Resilience Goals

Resilience engineering provides a conceptual framework for designing systems that can cope with disturbances, uncertainties and unforeseen situations. We have defined resilience [1, 2] as follows:

*A CPS (oS) is resilient if it has the ability to react to specified and unspecified disturbances in a way that preserves its function and reacts quickly. This reaction includes the early detection, minimization, prediction or even avoidance of disturbances. In addition it needs to have the capability to anticipate future challenges and to prepare itself for those".*

This definition focuses on the following resilience goals:

- **Anticipation:** Identifying risks early on, before they become critical.
- **Resistance:** Withstanding disturbances in order to protect critical functions.
- **Recovery:** Quickly restoring functionality after disturbances.
- **Error analysis:** Systematically identifying the causes of disturbances and malfunctions to prevent them from recurring.
- **Adaptation:** Dynamically adapting systems and organisations to new conditions and expanding their resilience potential in the long term.

These goals are particularly central to the cyber resilience life cycle [2] of a CPSoS, as these systems are operated over long periods of time and must cope with constantly changing environmental conditions. Based on these resilience goals, resilience architectures [10–16] have been developed in various standards, frameworks and research projects that translate these objectives into concrete levels and measures.

### 2.2 Overview of Established Resilience Architectures

Existing frameworks and models for resilience in technical systems [10–16] can be roughly grouped into four levels. For the synthesis underlying Table 1, we analysed a set of widely cited resilience and risk management frameworks that cover both socio-technical and cyber-physical aspects [10–16]. These include, among others, Hollnagel's Safety-II and Resilience Engineering concepts [12], the ISO 31010 risk management standards [15], NIST SP 800-160 Vol. 2 [11], ENISA cybersecurity guidelines [14], and recent architecture- and KPI-oriented approaches for CPS and CPSoS [10, 13, 16]. They were selected because they

(i) explicitly address resilience as a system property,
(ii) are either widely adopted standards or highly cited conceptual frameworks, and
(iii) together span technical, organisational and governance layers.

Table 1 summarises the four levels derived from this analysis. It assigns each level a specific primary goal and typical elements or measures, such as redundancy, Intrusion Detection Systems (IDSs), emergency response plans and risk assessments. In addition, the column 'Treatment of Adaptation' illustrates that adaptation is only considered implicitly in existing frameworks, rather than being implemented as an explicit, operational level.

*Table 1: The table summarises the layers of resilience architectures established in the literature. Each level is assigned to a specific objective and is operationalised by typical measures ranging from technical robustness to strategic governance. While universal countermeasures are primarily located at the structural level and specific countermeasures are implemented at the risk-specific level, the process and learning levels as well as the management level primarily address reactive or periodic aspects. The column 'Treatment of Adaptation' illustrates that the resilience goal of Adaptation is only implicitly considered in existing frameworks – either as a reactive follow-up activity after incidents (e.g., incident reviews) or as a periodic governance measure (e.g., risk assessments). An explicit, continuous, and operational implementation of Adaptation as a separate architectural level is lacking.*

| No. | Layer | Goal | Typical Elements / Measures | Treatment of Adaptation |
|---|---|---|---|---|
| 1. | **Structural Layer** | Baseline robustness | • Redundancy<br>• Diversity<br>• Failover<br>• Hardening | None – static by design; no self-adjustment |
| 2. | **Risk-Specific Defense Layer** | Targeted risk mitigation | • Firewalls<br>• IDS/IPS[1]<br>• Fire protection<br>• Targeted patching | None – static; activation only after predefined triggers |
| 3. | **Process and Learning Layer** (precursor of the Adaptation & Learning Layer (AL) in the proposed architecture) | Reaction and improvement after incidents | • Incident response<br>• post-incident reviews<br>• emergency plans<br>• training | Implicit – reactive |
| 4. | **Integration & Steering Layer** (management and governance functions, periodically updated) | Framework for governance and strategy | • Policies<br>• Risk Assessments<br>• Periodic audits | Implicit – periodic reviews and manual adjustments |

Essentially, existing models focus on two key areas:

- **Universal countermeasures** (structural layer: e.g. redundancy, segmentation, hardening), which have a system-wide effect and are located at the structural level.
- **Specific countermeasures** (risk-specific defense layer: e.g. firewalls, fire protection, malware detection), which are tailored to specific risks and located at the risk-specific level.

In addition, most frameworks include a process and learning layer that reacts to disturbances (e.g. incident response, post-incident reviews, training). This is supplemented by an integration and steering layer that formulates strategic guidelines (e.g. policies, audits, risk assessments).

---

[1] Intrusion Prevention System

Although Adaptation is identified in the literature as a key resilience goal, it is not implemented as an explicit operational layer in existing frameworks [10–16]. Instead, Adaptation only appears implicitly – for example, in the form of post-incident reviews or periodic governance processes. This means that it remains reactive rather than continuous.

A closer look reveals that:

- The Process and Learning Layer is not designed as a continuous, operational layer, but acts predominantly downstream. It lacks direct access to the ongoing coordination of countermeasures, meaning it can only act reactively.
- The Integration & Steering Layer primarily provides governance elements such as policies, standards and risk assessments. These activities are carried out periodically and not in real time. Accordingly, this layer cannot make immediate operational decisions, such as activating or deactivating specific countermeasures.

This means that all existing architectures lack a continuous operational control layer that coordinates risks, forecasts and countermeasures in real time. This *'adaptation gap'* forms the starting point of our approach: the introduction of an Adaptive Coordination Layer (ACL) that closes this missing bridge.

In addition to introducing an Adaptive Coordination Layer, this gap also requires a redefinition of the process and learning layer. In our approach, a downstream, reactive layer becomes an operational Adaptation & Learning Layer that continuously integrates experience, adapts policies and thus ensures the long-term development of resilience capabilities.

## 3. Related Work

As described in section 2.2, the majority of existing approaches in resilience engineering follow a two-group structure of countermeasures. Table 2 shows examples of how this classification is implemented in the literature. Universal measures include system-wide and long-term mechanisms such as redundancy, segmentation, or continuous monitoring. Specific measures, on the other hand, address concrete threats, for example through firewalls, fire protection, or tailor-made incident response playbooks.

*Table 2: The table shows selected works that address universal and specific countermeasures in resilience architectures. Universal measures such as redundancy, diversity, or monitoring have a system-wide effect, while specific measures target individual threats, for example through patching or intrusion detection techniques. The overview exemplifies that this two-group logic is widespread in the literature.*

| Source | Universal Countermeasures (System-wide) | Specific Countermeasures (Risk-targeted) |
|---|---|---|
| Woods (2015) – **Four Concepts for Resilience** [17] | • Monitoring<br>• Anomaly Detection<br>• Loose Coupling of Components | • Training of operating personnel for critical processes |
| Kott et al. (2018) – **Approaches to Enhancing Cyber Resilience** [18] | Generic strategies such as<br>• Dynamic defence (moving target defence)<br>• Continuous monitoring<br>• Redundancy principles<br>• Recovery mechanisms | Threat-specific measures for individual attack vectors, e.g.<br>• Targeted malware defence<br>• Attack modelling and simulation<br>• Customized intrusion detection |
| **ISO 31010:2019** (2019) – **Risk Management, Risk Assessment Techniques** [15] | • Redundancy in system design<br>• Regular maintenance intervals | • Fire protection measures in buildings<br>• Firewalls for cyber risks |

| NIST SP 800-160 Vol. 2 (2021) – **Developing Cyber Resilient Systems [10]** | • System segmentation<br>• Redundancy<br>• Diversity<br>• Continuous monitoring | • Malware-specific Detection<br>• Tailored Access Controls<br>• Incident-specific Response Playbooks |
|---|---|---|
| Hollnagel (2018) – **Safety-II & Resilience Engineering [12]** | • Increasing robustness through modular architectures | • Specialized sensors for overheating risks |
| Berger et al. (2021) – **A Survey on Resilience in the IoT: Taxonomy, Classification and Discussion of Resilience Mechanisms [19]** | Principles such as<br>• Redundancy<br>• Network diversity<br>• General system monitoring | Measures targeting specific device groups, e.g.<br>• Targeted recovery<br>• Fault tolerance for certain sensor subsystems |
| **Jalowski et al. (2022)** – **A Survey on Moving Target Defense** (MTD) **for Networks: A Practical View [20]** | General MTD strategies such as<br>• Distributed IP rotation<br>• Topology changes<br>• Network diversity | • MTD measures tailored to specific attack scenarios or defences against particular exploits |
| **ENISA (2024)** – **Cybersecurity Guidelines for Critical Infrastructure [14]** | • Zero-trust architecture<br>• Network segmentation | • Threat-specific Vulnerability Patching,<br>• Targeted Security Drills |
| Salayma (2024) – **Risk and threat mitigation techniques in internet of things (IoT) environments: a survey [21]** | • Redundancy in multi-sensor technologies, network reconfiguration to maintain integrity | • Isolation of compromised devices<br>• Selective blocking of subnets<br>• Specific patching for identified vulnerabilities |
| Rauti et al. (2024) – **Enhancing resilience in IoT cybersecurity: the roles of obfuscation and diversification techniques for improving the multilayered cybersecurity of IoT systems [22]** | • Diversification of system architectures<br>• Multi-layer obfuscation | • Obfuscation of specific interfaces<br>• targeted diversification for vulnerable components |

This overview illustrates that the two-group paradigm is widely represented in the literature – from classic concepts such as the 'Four Concepts for Resilience' (Woods 2015) [17] to international standards (ISO 31010 [15], NIST SP 800-160 [10]) and practice-oriented guidelines (ENISA 2024) [14]. More recent works such as Berger et al. (2021) [19], Jalowski et al. (2022) [20], Salayma (2024) [21] and Rauti et al. (2024) [22] expand this picture in the context of the Internet of Things and dynamic network security. Nevertheless, the treatment of adaptation remains limited at all levels. Woods [17] emphasises 'graceful extensibility' as the ability to respond beyond system boundaries, but does not anchor this in an architectural level. NIST SP 800-160 [10] mentions 'adapt' as a resilience goal without making it operational. ENISA [14] focuses on crisis management and ad hoc responses, not on continuous control.

Adaptation is also discussed in other domains such as energy supply, aviation and healthcare, but mostly as part of organisational learning or safety culture (e.g. Hollnagel (2018) [12] in the context of Safety II). These contributions make it clear that although adaptation is recognised in theory, in practice it is often only understood as a downstream process.

Only recently have data-driven methods opened up new possibilities. IoT monitoring, machine learning and KPI (key performance indicator) tracking make it possible to dynamically record and continuously evaluate resilience metrics [20, 21]. Nevertheless, there has been no systematic integration of these metrics into an architectural control level to date. Existing work operates predominantly at the level of individual techniques or selective mechanisms without embedding them in a systematic architectural framework.

Despite their widespread use, the various approaches have key limitations:

- Countermeasures are predominantly static and essentially only reduce risks at the time they are introduced.
- Adaptation and continuous learning are not explicitly taken into account, but appear at most implicitly in downstream processes such as post-incident reviews or periodic policy updates.
- There is no feedback loop between risk assessment, the effectiveness of measures and their active adjustment.

This analysis highlights that although existing work [10, 12, 14, 15, 17–22] consistently addresses universal and specific countermeasures, it does not provide an explicit operational level for continuous adaptation. This is precisely where the approach proposed in this paper comes in, with the introduction of an Adaptive Coordination Layer (ACL) in combination with a redefined Adaptation & Learning Layer.

## 4. Proposed Concept: Adaptive Coordination Layer & Adaptation Layer

The Adaptive Coordination Layer (ACL) is the core operational component of the proposed resilience architecture. It acts as an executive layer that not only triggers existing countermeasures, but also dynamically controls their activation, prioritisation and parameterisation depending on the current system status. In this way, the ACL closes the gap that previously existed between analytical risk detection and strategic adaptation and, together with the redefined Adaptation & Learning Layer (AL), forms an integrated mechanism for short-term response and long-term adaptation. While the ACL is responsible for operational implementation, the AL takes over the ongoing evaluation and optimisation of decision-making logic based on past events.

Table 3 presents the extended resilience architecture, featuring the newly introduced Adaptive Coordination Layer (orange, Layer 5) and the newly defined Adaptation & Learning Layer (orange, Layer 3). Both layers complement the existing structure by adding operational and strategic dimensions of adaptation.

*Table 3: The table summarises a multi-layer resilience architecture, linking structural robustness, risk-specific defences, governance, and adaptive coordination. While lower layers (4–5) rely on static mechanisms, higher layers (1–3) progressively incorporate feedback, learning, and real-time adaptation through the Adaptive Coordination Layer (ACL) and continuous data flows across all layers.*

| No. | Layer | Goal | Typical Elements / Measures | Treatment of Adaptation |
|---|---|---|---|---|
| 1. | **Adaptive Coordination Layer (ACL) (new)** | Meta-strategic coordination and prioritisation – system-wide balance of countermeasures, resource allocation, and initiation of learning processes | • Strategy shifts (e.g. from cyber defence to physical robustness) <br>• Coordination of universal and specific countermeasures <br>• Escalation to the Adaptation Layer | Explicit, real-time adaptation based on KPI deviations and system state changes; continuous feedback loop with AL. |
| 2. | **Integration & Steering Layer** | Framework for governance and strategy | • Policies <br>• Risk Assessments <br>• Periodic audits | Implicit – periodic reviews and manual adjustments |
| 3. | **Adaptation & Learning Layer (re-defined)** | Long-term improvement and policy adjustment | • Post-incident reviews <br>• Policy updates <br>• Training programs <br>• Threshold adjustments | Continuous learning based on feedback from ACL and control layer; operationalizes strategic adaptation within the system lifecycle. |
| 4. | **Risk-Specific Defense Layer** | Targeted risk mitigation | • Firewalls <br>• IDS/IPS <br>• Fire protection <br>• Targeted patching | None – static; activation only after predefined triggers |
| 5. | **Structural Layer** | Baseline robustness | • Redundancy <br>• Diversity <br>• Failover <br>• Hardening | None – static by design; no self-adjustment |
| hor.[2] | **Data & Information Flow (cross cutting layer)** | Continuous monitoring, prediction, and feedback loops | • Sensor systems <br>• KPI dashboards <br>• Log analysis <br>• AI/ML models <br>• Data integration | Enables adaptive information exchange and cross-layer learning support. |

Table 3 illustrates the expanded resilience architecture with five hierarchically and functionally differentiated levels. At the top is the ACL, which handles the operational coordination and prioritisation of countermeasures. The ACL forms the interface between analytical assessment and active control. It coordinates countermeasures across physical, cybernetic and organisational subsystems, thus ensuring a uniform response to dynamically changing disruption and stress situations. Its decisions are based on continuously incoming status data, key performance indicators (KPIs) and forecasts from the analysis and monitoring levels. On this basis, the ACL selects appropriate measures, adjusts their intensity and coordinates parallel interactions between subsystems in near real time. The focus is thus on short-term, operational adaptation.

---

[2] Horizontal Layer (Cross-cutting Layer)

Below the ACL lies the Control Layer, which integrates data aggregation, system state analysis, and forecasting models (e.g. KPI tracking or machine learning).

The subsequent AL functions as an evaluation and optimisation layer. It analyses feedback from the ACL, refines decision rules, adjusts thresholds, and feeds the acquired knowledge back into ongoing operations. The AL thus focuses on long-term structural optimisation.

Building on this, the Risk-Specific Defense Layer implements targeted countermeasures, while the Structural Layer provides fundamental protection and redundancy mechanisms that operate independently of specific disturbance scenarios.

Together, the ACL and AL form a closed adaptation cycle that extends classical resilience architectures with a continuous feedback and learning mechanism. This transforms the adaptation logic across all layers — from reactive and static to proactive, continuous, and system-immanent. Adaptation is no longer understood as a consequence of events but as a permanent system capability.

Figure 1 illustrates the functional embedding of the ACL in the extended resilience architecture and highlights its key role as a coordination and mediation layer that integrates the vertical information and control processes between technical, organisational and learning-based levels.

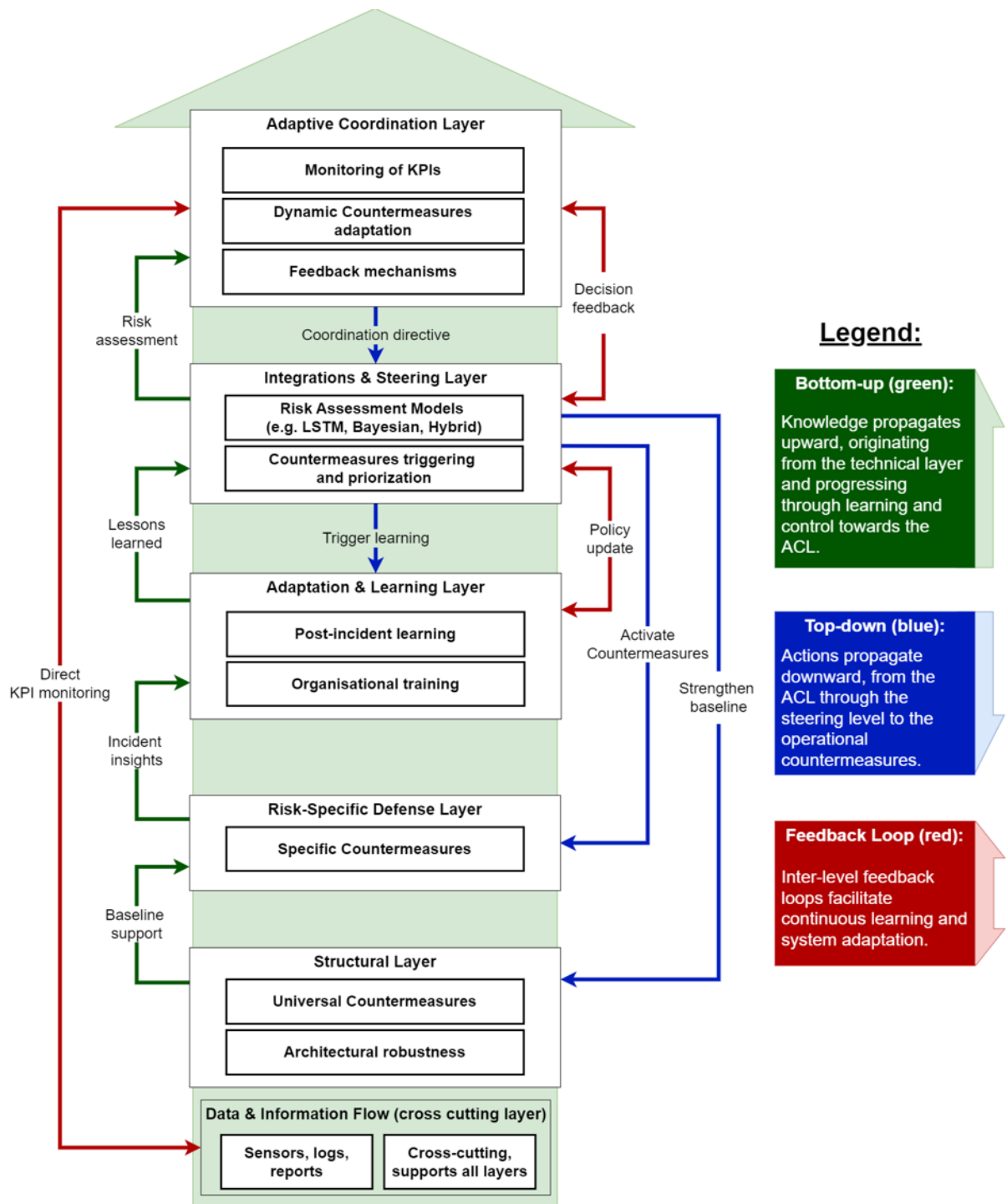

Figure 1: Functional integration of the ACL within the extended resilience architecture. The architecture consists of six interconnected layers, ranging from the Data & Information Flow Layer at the base to the Adaptive Coordination Layer at the top.
Knowledge and insights propagate bottom-up (green arrows) from technical layers through learning and steering levels towards the ACL. Conversely, actions and control directives propagate top-down (blue arrows), originating from the ACL and cascading through the steering, learning, and defense layers to the structural baseline.
Feedback loops (red arrows) interconnect all layers, enabling continuous monitoring, evaluation, and adaptation. The ACL plays a central role in monitoring KPIs, dynamically adapting countermeasures, and providing feedback to ensure coordinated, system-wide resilience across technical, organisational, and learning domains.

Figure 1 illustrates the vertical integration of technical (Structural Layer and Risk-Specific Defenses), organisational (Integration & Steering Layer and ACL), and learning-oriented levels (Adaptation & Learning Process and its feedback with the ACL), as well as their bidirectional coupling through information, decision, and feedback flows. The flows are detailed in Table 4, which distinguishes between information and knowledge flows (bottom-up, green), decision flows (top-down, blue), and feedback flows (red).

*Table 4: The following description details the three interrelated flow types (bottom-up, top-down, and feedback) illustrated in Figure 1. Each flow connects specific layers within the architecture and supports continuous adaptation through information exchange, decision propagation, and performance feedback.*

**Information and Knowledge Flow (Bottom-up, green):**

- **Structural Layer → Risk-Specific Defense Layer: Baseline support**
  **Meaning:** The robust baseline architecture (e.g. redundancy) reinforces the effectiveness of specific countermeasures.
  **Example:** A redundant backup system (a universal countermeasure at the structural level) ensures that data are not lost even in the event of a failure or security incident. This allows the firewall-specific countermeasure (a risk-specific measure, e.g. blocking network traffic) to be applied more aggressively or safely — for instance, by terminating connections more quickly without endangering operations.
- **Risk-Specific Defense Layer → Adaptation & Learning Process: Incident insights**
  **Meaning:** Specific defense measures provide feedback insights ("Did the attack succeed? Was the countermeasure effective?").
  **Example:** An antivirus system detects malware → the result (effective or not) is reported to the learning process.
- **Adaptation & Learning Process → Integration & Steering Layer: Lessons learned**
  **Meaning:** Lessons derived from incidents provide updated rules or improvements for the control system.
  **Example:** After evaluation, the system reports: "Firewall signatures need to be updated earlier."
- **Integration & Steering Layer → ACL: Risk assessment**
  **Meaning:** The Integration & Steering Layer provides the ACL with risk analyses, KPI monitoring results, and model outputs.
  **Example:** "Probability of attack X = 70%" → the ACL uses this information for coordination decisions.

**Decision Flow (Top-down, blue):**

- **ACL → Integration & Steering Layer: Coordination directive**
  **Meaning:** The ACL makes strategic decisions ("Which countermeasures should be activated?") and issues commands to the Integration & Steering Layer.
  **Example:** The ACL instructs: "Prioritise cyber defence over physical security for the next 30 minutes."
- **Integration & Steering Layer → Adaptation & Learning Process: Trigger learning**
  **Meaning:** The Integration & Steering Layer triggers targeted learning when new patterns or emerging risks are detected.
  **Example:** "This attack pattern is new — forward it to the Adaptation & Learning Layer."
- **Integration & Steering Layer → Risk-Specific Defense Layer: Activate countermeasures**
  **Meaning:** The Integration & Steering Layer activates specific defense mechanisms.
  **Example:** "Activate Intrusion Detection Rule 5."
- **Integration & Steering Layer → Structural Layer: Strengthen baseline**
  **Meaning:** The Integration & Steering Layer reinforces the structural baseline when necessary.
  **Example:** "Bring redundant server online."

> **Feedback Flows (red):**
> - **ACL ↔ Integration & Steering Layer: Decision feedback**
>   **Meaning:** The Integration & Steering Layer reports back on how decisions were implemented; the ACL adjusts its strategies accordingly.
>   **Example:** ACL: "Activate redundancy." → Integration & Steering Layer: "Completed, effectiveness 90%."
> - **Integration & Steering Layer ↔ Adaptation & Learning Process: Policy update**
>   **Meaning:** The Integration & Steering Layer provides data on countermeasure effectiveness → the Adaptation & Learning Process updates policies and returns refined rules.
>   **Example:** "Recovery took 20 minutes instead of 10 → update recovery policy."
> - **ACL ↔ Data & Information Flow (cross-cutting layer): Direct KPI monitoring**
>   **Meaning:** The ACL can directly access system metrics (e.g. system health) without going through the Integration & Steering Layer.
>   **Example:** The ACL detects that CPU utilisation exceeds 90% → orders resource reprioritisation.

In summary, this means that knowledge rises from the operational layers to the ACL via structured learning processes (information or knowledge flow, green), while strategic decisions, control commands and prioritisations are passed down to the lower levels (decision flow, blue). In addition, the data & information flow layer runs horizontally across all levels and ensures the continuous exchange of data and knowledge, enabling adaptive control and continuous learning (feedback flows, red).

The ACL acts as the operational control instance, receiving risk assessments, status data and KPIs from the analysis levels and using them to derive dynamic adjustments to ongoing countermeasures. The feedback loops continuously link all levels with each other, creating a closed learning and adaptation cycle. This interaction enables real-time coordination between the physical, cybernetic and organisational levels, thereby operationalising the resilience objective of adaptation as a permanent system capability.

While the lower layers of the architecture implement specific countermeasures to reduce risk, the ACL assumes a coordinating and prioritising role. To illustrate this functional position of the ACL within the overall architecture, Table 5 differentiates it from the existing categories of universal and specific countermeasures.

*Table 5: The table contrasts the functional characteristics of traditional countermeasures (universal and specific) with those of the Adaptive Coordination Layer (ACL). While universal and specific countermeasures operate at the tactical level and focus on mitigating individual risks, the ACL functions at a meta-level as a dynamic coordination instance. It continuously balances trade-offs between different countermeasures, prioritises resource allocation, and adapts decisions based on real-time data, KPIs, and predictive models.*

| Aspect | Universal / Specific CMs | Adaptive Coordination Layer (ACL) |
|---|---|---|
| **Role** | Implementation of measures to reduce risk or mitigate impact | Control and decision-making layer |
| **Level of action** | Executing (tactical layer) | Coordinating (meta-level) |
| **Temporal reference** | Static (universal) / reactive (specific) | Dynamic and continuous |
| **Decision logic** | Defined by policies or fixed rules | Data-driven and adaptive (e.g. based on KPIs, forecasts, or ML models) |
| **Objective** | Acts on individual risks or subsystems | Balances the overall system (trade-offs between CMs, resource allocation, and prioritisation) |

| **Example** | Firewall blocks an attack | ACL decides whether cyber defence or physical redundancy should be prioritised |

In existing frameworks, countermeasures are usually considered as isolated elements – either universal (permanently effective) or specific (risk-related). However, there is no explicit control logic that dynamically coordinates these measures.

The ACL addresses precisely this gap: it functions as a meta-strategic level that dynamically adjusts the effectiveness, sequence and priority of countermeasures depending on the current system status. This fundamentally distinguishes the ACL from classic countermeasures: it does not execute measures, but decides on their optimal combination and timing of activation. This separation is conceptually crucial for developing resilience from a sum of isolated protection mechanisms into a coordinated, adaptive system capability.

In the following section, five different implementation variants of the ACL are described, which differ in terms of decision logic, complexity and degree of automation. Each variant represents a specific development approach for implementing operational adaptation in CPSoS architectures and gradually transitioning from rule-based to self-learning coordination mechanisms.

## 5. Implementation Approaches

Following the conceptual and architectural classification of ACL and AL, this section describes potential approaches for practical implementation. The aim is to demonstrate the fundamental feasibility of the proposed layers and to highlight different implementation paths that vary in terms of technical complexity, degree of automation and integration effort. The variants presented are not alternatives, but rather different degrees of maturity on the path to fully adaptive resilience mechanisms.

### 5.1 Adaptive Coordination Layer (ACL)

A uniform reference scenario is used to improve the comparability of the following implementation variants. It describes a situation in which an increased probability of failure of a critical sensor or subsystem is detected within a CPSoS. This malfunction poses a risk of functional impairment at the system level and requires coordinated adjustment of countermeasures.

In this scenario, the ACL decides how universal and risk-specific CMs should be combined, prioritised and parameterised in order to maintain system stability. Depending on the technical characteristics of the ACL – whether rule-based, KPI-driven, learning-based or distributed – the way in which information is processed, decisions are made and feedback is used differs.

This consistent application scenario enables a direct comparison of the functional principles, strengths and limitations of the individual variants and serves as a consistent framework for evaluating operational adaptability.

The different characteristics of the ACL can be evaluated on the basis of several dimensions:

- **Complexity:** This dimension describes the structural and algorithmic effort required to implement the decision logic. Higher complexity generally enables greater adaptability, but at the same time increases the requirements for computing power, integration and maintenance.

- **Transparency:** This refers to the traceability and explainability of the decisions made within the ACL. Variants with high transparency are easier to audit and validate, while learning-based systems often have lower interpretability.
- **Proactivity:** This dimension evaluates the ACL's ability not only to detect deviations, but also to anticipate future state changes at an early stage. A high level of proactivity contributes to reducing response times and preventive risk mitigation.
- **Implementation effort:** This describes the technical and organisational effort required to integrate the respective variant into existing CPSoS architectures. This includes requirements for data availability, model training, interfaces and IT infrastructure.

These dimensions constitute the evaluation framework used to describe and compare the following implementation variants.

**Variant 1 – Rule-Based Policy Engine (simple, deterministic, transparent)**

This variant represents the classical starting point for adaptive coordination. It is based on fixed decision rules that are predefined and maintained manually. The ACL essentially acts as a rule-driven orchestration layer that maps simple *if–then* relationships between system states and countermeasures.

The focus lies on clear reproducibility and high explainability of decisions – ideal for safety-critical or regulated environments, where traceability takes precedence over autonomy.

- **Core idea:** Adaptation is achieved through deterministic *if–then* rules linking specific states to predefined countermeasures.
- **Technical implementation:** Policy engines or decision trees evaluate sensor data, risk probabilities, and KPI trends.
- **Example:** If the probability of failure of a temperature sensor exceeds 0.7, the ACL activates a redundant sensor unit. If the temperature deviation is greater than 5 K, a cooling CM with higher priority is triggered. If the mean time to recovery surpasses the target value, the ACL automatically notifies the maintenance system.
- **Reaction mode:** Decisions are made strictly on a rule-based foundation, without context evaluation or learning components.

This variant is characterised by low complexity and maximum transparency. The decision rules are clearly defined, easy to understand, and auditable. However, its proactivity is very limited, as it only reacts to already detected deviations. The implementation effort remains minimal, as neither learning models nor continuous data integration are required.

This approach is therefore particularly suitable for infrastructure-oriented CPSoS, such as energy or railway systems, where stability and traceability are prioritised over adaptive intelligence. In these environments, operational processes are highly regulated, and safety requirements demand fully verifiable decision logic. As a result, deterministic rule-based mechanisms are often preferred, since they allow for transparent validation and ensure predictable system responses under all operating conditions.

**Variant 2 – KPI-based feedback module (metric-driven adjustment)**

This variant represents the first evolutionary extension of rule-based control. Instead of checking only predefined states, the ACL continuously monitors key performance indicators (KPIs) that reflect the system state in real time. Deviations from target values are automatically detected and used to control countermeasures. This creates a data-based feedback loop that marks the transition from reactive to dynamic adaptive control.

- **Core idea:** The ACL responds to quantitative changes in KPIs such as sensor availability, average temperature deviation or time-to-mitigation.
- **Technical implementation:** Threshold analyses, moving averages and anomaly detection are combined to identify trends and deviations.
- **Example:** If sensor availability falls below 95% or the temperature deviation rises above 5 K, the ACL reduces the sampling rate of other, less critical sensors in order to redirect resources. If increased fluctuation is detected, the ACL activates a universal cooling CM to reduce the load.
- **Reaction mode:** Decisions are data-driven and iterative; adjustments are continuously evaluated and fine-tuned.

The KPI feedback module has a moderate level of complexity, as it already integrates data aggregation and threshold logic. Transparency remains high: all decisions can be traced via measurable KPI deviations. Proactivity is slightly higher than in the rule-based variant, as trends can be identified at an early stage before thresholds are violated. The implementation effort is moderate and depends heavily on the existing data quality and infrastructure. This variant thus offers a good compromise between comprehensibility and data-driven adaptivity.

The use of this approach is particularly advantageous in production-related or logistical CPSoS, where performance indicators can be continuously monitored and operational adjustments can be made based on well-defined target values. In such systems, stability is closely linked to measurable performance metrics, which makes KPI-driven adaptation both practical and effective without introducing excessive algorithmic complexity.

**Variant 3 – Machine learning-based decision agent (learning, adaptive, data-driven)**

This variant represents the transition to autonomous, experience-based control approaches. Instead of responding to fixed thresholds, the ACL uses machine learning to independently identify optimal combinations of measures. It acts as a learning agent that combines historical patterns, simulation results and current sensor data to make decisions that maximise system resilience in the long term.

- **Core idea:** The agent learns correlations between input parameters (e.g. temperature, vibration, failure probability) and success metrics (e.g. MTTR, risk reduction).
- **Technical implementation:** Reinforcement learning algorithms or classification models are trained to develop strategies that minimise risk and ensure system stability.
- **Example:** The agent recognises that early activation of redundancy sensors, combined with a 2% increase in cooling capacity, produces more stable temperature curves and lower overall costs in the long term than a reactive restart.
- **Reaction mode:** Decisions are made based on continuous learning and evaluation of previous actions; the agent simulates different strategies and selects the most effective one.

This variant leads to significantly higher complexity, as models must be trained, validated and continuously optimised. Transparency decreases because decision-making logic in learning models can only be explained to a limited extent. On the other hand, proactivity increases significantly: the agent can predict future failures and take preventive action instead of merely reacting to events. The implementation effort is high, particularly due to the need for data, computing power and model maintenance. Overall, this variant offers high adaptation potential, but requires careful monitoring and continuous model evaluation.

This approach is particularly suitable for data-intensive and dynamic CPSoS, such as industrial manufacturing networks or smart energy grids, where non-linear dependencies and time-varying risk dynamics must be captured. In such environments, learning-based agents provide the necessary flexibility to handle complex interactions and evolving operational conditions.

**Variant 4 – LLM-supported decision logic (hybrid AI assistance with contextual understanding)**

This variant extends the ACL into the domain of semantic and context-aware decision support. In addition to structured sensor data, unstructured textual information from maintenance logs, fault reports, or system records are also considered. An integrated large language model (LLM) expands the ACL's decision-making basis by recognising relationships that are not apparent in numerical data. This results in a hybrid architecture combining deterministic logic with generative AI, making use of both contextual knowledge and quantitative evidence.

- **Core idea:** The LLM analyses textual data to identify patterns or cause-and-effect relationships and derives recommendations for corresponding adjustments.
- **Technical implementation:** The LLM is connected via a semantic interface; its suggestions are verified and validated by the rule- and KPI-based modules.
- **Example:** The LLM detects in log entries that sensor failures frequently occur under high humidity conditions and recommends temporarily lowering the humidity threshold and activating a preventive cooling mode. The ACL checks this suggestion against current KPIs and implements it if deemed plausible.
- **Reaction mode:** A combination of rule-based and context-sensitive decision processes that integrate qualitative empirical information.

LLM-supported decision logic is characterised by high complexity and a correspondingly large implementation effort, as both semantic models and validation mechanisms must be integrated. Transparency remains limited, since generated recommendations are often difficult to interpret in full. Nevertheless, this variant achieves a high degree of proactivity by incorporating contextual information from unstructured data sources and recognising causal relationships that purely numerical models would not capture.

This approach is particularly suitable for systems with substantial knowledge requirements, where context sensitivity is more important than full explainability. It finds practical application in knowledge-intensive CPSoS such as fleet-management or maintenance networks, where unstructured information from logs, reports, or communication records can be incorporated directly into the decision-making process.

**Variant 5 – Autonomous multi-agent architecture (distributed, scalable, cooperative)**

This variant represents the most advanced form of ACL. It transfers the decision-making logic to a network of specialised software agents, each of which autonomously handles specific tasks. The ACL takes on the coordination and prioritisation function between these agents, creating distributed, real-time resilience control. Such architectures are particularly suitable for large-scale CPSoS, in which individual subsystems make local decisions but must remain synchronised globally.

- **Core idea:** Multiple specialised agents (e.g. *SensorAgent*, *MaintenanceAgent*, *CoolingAgent*) interact via standardised communication protocols to coordinate local decisions and avoid conflicts.
- **Technical implementation:** Frameworks such as JADE [23] or ROS2 [24] enable agent-based communication via a central message bus, which is monitored and coordinated by the ACL.

- **Example:** The *SensorAgent* reports an increasing probability of failure; the *MaintenanceAgent* initiates preventive calibration, while the *CoolingAgent* adjusts the cooling capacity in parallel. The ACL evaluates the overall impact, prioritises the measures and coordinates their execution system-wide.
- **Reaction mode:** Decisions are made decentrally and coordinated in real time; the ACL ensures consistency and system balance.

The multi-agent architecture is the most complex of all variants, as multiple decentralised units must be coordinated and synchronised. Transparency is limited because decisions are made in a distributed manner and can only be traced at the system level. Proactivity reaches its maximum: agents can react locally and collaborate globally to prevent system failures before they escalate. The implementation effort is very high – in terms of integration, communication and governance.

This variant is especially suitable for large-scale CPSoS that require robust, scalable, and largely autonomous adaptation. It is particularly effective in highly distributed environments such as smart cities, traffic control systems, or interconnected production sites, where local agents operate autonomously while the ACL maintains overall coordination.

The five ACL variants represent different manifestations of the resilience principle of adaptation, each with distinct advantages depending on the system context. The rule-based policy engine (Variant 1) provides a deterministic and auditable form of controlled adaptation, suitable for highly regulated CPSoS such as energy or railway systems. The KPI-based feedback module (Variant 2) builds upon this stability by adding data-driven feedback loops that enable precise responses to measurable deviations. Machine learning-based decision logic (Variant 3) offers greater contextual sensitivity, allowing complex relationships and dependencies to be considered in real time. The LLM-supported decision logic (Variant 4) extends this concept through semantic analysis and the integration of unstructured information, for instance from maintenance logs. Finally, the autonomous multi-agent architecture (Variant 5) enables a decentralised and self-organising form of coordination, particularly suited for highly dynamic, networked CPSoS.

These variants should not be viewed as a linear development sequence but rather as a spectrum of adaptive control approaches that can be applied differently depending on the required balance between complexity, safety, and transparency. Resilience is therefore not understood as a hierarchically progressive capability but as a context-dependent and configurable property. The appropriate level of implementation can be selected according to system maturity, data availability, and the trade-off between transparency and autonomy. Together, these approaches constitute the operational core of an adaptive resilience architecture.

However, the five variants differ not only in their technical implementation but also with respect to key evaluation criteria such as complexity, transparency, proactivity, and implementation effort. Table 6 summarises their main characteristics, strengths, and limitations and illustrates the gradual transition from deterministic to autonomous, adaptive control mechanisms.

*Table 6: The Table summarises the main characteristics of the five ACL variants. Each variant represents a specific approach to adaptive coordination within CPSoS, differing in terms of decision logic, data processing, and implementation complexity. The table highlights the core principles, decision mechanisms, and exemplary reactions, as well as the respective strengths and limitations with regard to transparency, proactivity, complexity, and implementation effort. The progression from rule-*

*based to learning-based and distributed architectures illustrates the transition from deterministic control mechanisms to increasingly autonomous and context-aware resilience strategies.*

| Variant | Core principle | Decision logic | Example reaction | Strengths (Transparency / Proactivity) | Limitations (Complexity / Implementation effort) |
|---|---|---|---|---|---|
| 1 – Policy Engine | Fixed rules (*if–then*) | Deterministic | Activates redundant sensor when $p >\backslash 0.7$ | • High transparency<br>• Easy traceability | • No learning capability<br>• Limited adaptability |
| 2 – KPI Feedback | Threshold monitoring | KPI-based | Adjustment triggered by KPI violation | • Quantifiable<br>• Clear triggering mechanism | • Purely reactive<br>• moderate implementation complexity |
| 3 – ML Agent | Machine learning | Learning, adaptive | Early activation of redundant sensor for prevention | • Anticipatory<br>• Identifies non-linear patterns | • High training effort<br>• Limited explainability |
| 4 – LLM Logic | NLP [3]+ rule engine | Context-based | Adjustment of maintenance intervals based on text analysis | • Context-sensitive<br>• Leverages unstructured data | • Validation required<br>• Bias risk<br>• Higher integration effort |
| 5 – Multi-Agent | Coordination of autonomous agents | Decentralised | Synchronous control of countermeasures across subsystems | • High proactivity<br>• Scalability<br>• robustness | • Complex synchronisation<br>• High communication overhead |

As Table 6 shows, both complexity and implementation effort increase with the growing autonomy of the ACL variants, while transparency decreases and proactivity rises significantly. The ACL thus forms a continuous line of development—from rule-based, easily comprehensible systems to data-driven and context-sensitive architectures.

However, this increasing autonomy at the operational level necessitates a higher-level authority to evaluate decisions, systematise experience, and translate the resulting insights into new policies over time.
This interplay between short-term adaptation and long-term strategic learning is addressed in the following section, which introduces the Adaptation & Learning Layer (AL).

5.2 Adaptation & Learning Layer (AL)

In this paper, the Adaptation & Learning Layer (AL) represents a redefinition of the classic learning layer in resilience architectures.

Whereas earlier approaches [10–16] mostly understood the Adaptation layer as a reactive, downstream instance – responsible for analysing past disruptions and deriving organisational improvements – it is redesigned here as a strategic-cooperative layer. The AL continuously interacts with the ACL, evaluates its operational decisions, recognises recurring patterns and translates these into structural and long-term adjustments at the policy, governance or architecture level.

---
[3] NLP (Natural Language Processing) techniques are used to extract and interpret relevant information from unstructured textual data such as maintenance logs or incident reports.

As a result, learning is no longer understood as a retrospective process, but as a permanent, data-driven and cross-system mechanism that gradually optimises the resilience of the overall system.

The same reference scenario as in the previous section is used for illustration purposes: Within a CPSoS, a critical temperature sensor shows an increased probability of failure. While the ACL reacts in the short term – for example, by activating redundant sensors or adjusting the cooling capacity – the AL subsequently analyses the effectiveness of these measures, learns from the results and adjusts decision rules, thresholds or guidelines on this basis.

This creates a closed resilience cycle in which the AL forms the long-term memory of the system. Similar to the ACL, the characteristics of the AL differ in four key dimensions:

- **Time horizon:** This dimension describes whether learning processes are event-driven (reactive) or continuous (preventive). Short time horizons serve to correct errors after disruptions, while long-term approaches accumulate knowledge permanently and use it to proactively adapt guidelines and parameters.
- **Organisational integration:** This describes the extent to which learning processes are anchored within the organisation – from locally in individual subsystems to system-wide or cross-industry networks. A high level of integration increases collective learning ability, but requires coordinated communication and trust structures.
- **Depth of learning:** This dimension characterises the extent to which insights gained influence existing structures. Superficial learning leads to parameter adjustments, whereas deep learning leads to policy or process changes and thus to sustainable system transformation.
- **Degree of automation:** This describes the extent to which learning processes are manual, standardised or AI-supported. Higher automation enables faster response and continuous improvement, but reduces transparency and control.

These dimensions form the evaluation framework for the following implementation variants of AL and illustrate how learning in CPSoS can evolve from reactive reflection to continuous, self-optimising knowledge integration.

**Variant 1 – Post-incident reviews & workshops (reactive, established, organisational)**

This variant represents the traditional basis of organisational learning and focuses on the retrospective analysis of disruptive events. It is usually carried out manually by experts and serves to identify causes and derive preventive measures. The learning processes are event-oriented, strongly human-centred and organisationally anchored.

- **Core idea:** Learning takes place retrospectively on the basis of documented events, with the aim of avoiding similar mistakes in the future.
- **Technical implementation:** Use of standardised methods such as root cause analysis, incident reports and lessons learned meetings.
- **Example:** After a series of temperature sensor failures, it is determined that the sensor housing seal fails in high humidity conditions. The maintenance policy is then adjusted to include regular checks of this seal.
- **Reaction mode:** Findings are consolidated manually and transferred to maintenance or operating guidelines.

This variant is characterised by low complexity and high transparency, as decisions are documented in a comprehensible manner. However, it has no degree of automation and is limited to short-term, local learning outcomes. Its value lies in the consolidation of experiential knowledge, rather than in system-wide or data-based optimisation.

This variant is typically used in regulated, safety-critical CPSoS – such as in energy supply or railway technology – where follow-up and documentation of incidents are a mandatory part of the operating process.

**Variant 2 – Continuous organisational learning loops (systematic, process-integrated)**

This variant establishes learning as an integral part of the organisational process. It transforms reactive knowledge acquisition into a cyclical, systematically embedded feedback structure, in which experiences are regularly collected, evaluated and translated into operational standards.

- **Core idea:** Experiences and KPIs from ongoing operations are continuously analysed in order to iteratively improve processes and maintenance strategies.
- **Technical implementation:** Integration into existing process management systems (e.g. PDCA cycle [25], lean management [26] or Kaizen [27]) with automated review triggers.
- **Example:** The ACL regularly transmits KPI trends on sensor failure probability to the AL. If a recurring pattern is detected (e.g. seasonal failures), maintenance schedules are automatically adjusted.
- **Reaction mode:** Learning takes place cyclically and is data-based over several operating periods.

This variant has medium complexity and high transparency. It offers increased proactivity compared to reactive reviews, as trends can be identified at an early stage. The degree of automation remains moderate, as human review and management decisions are still required.

It is particularly suitable for organisations that want to integrate learning into their operations in a structured way without relying entirely on automation. A typical area of application is process-oriented CPSoS such as manufacturing or chemical plants, where continuous monitoring of key performance indicators enables the adjustment of thresholds and strategies.

**Variant 3 – Policy & governance update mechanisms (structured, standardised, verifiable)**

This variant transforms learning into a formal governance process. It links findings from ACL and AL analyses directly to policy and audit cycles in order to permanently embed structural adjustments. This institutionalises learning and makes it verifiable.

- **Core idea:** Results from operational decisions are reviewed at regular intervals and, if necessary, lead to adjustments to policies or security standards.
- **Technical implementation:** Automated reporting systems and audit tools based on established frameworks (e.g. ISO 27001, NIST Cybersecurity Framework).
- **Example:** Repeated sensor malfunctions under certain environmental conditions prompt the AL to reduce the permissible operating parameters for sensors in the policy document and define a new maintenance routine.
- **Reaction mode:** Adjustments are made according to plan in formal review cycles with documented approval and feedback.

This variant offers a high degree of transparency and governance compliance, as all changes are documented in a traceable manner. The complexity is moderate to high, as interfaces between operational activities and the governance structure must be coordinated. Proactivity depends on the review frequency, and the degree of automation remains limited.

This variant is ideal for regulated industries where verifiability and auditability are paramount. It is therefore particularly suitable for distributed maintenance or delivery networks where experiential knowledge can be systematically transferred into operational guidelines and control logic.

It is particularly suitable for organisations that want to integrate learning into their operational processes in a structured manner without relying entirely on automation. A typical area of application is process-oriented CPSoS such as manufacturing or chemical plants, where continuous monitoring of key performance indicators enables the adjustment of thresholds and strategies.

**Variant 4 – Data-driven adaptation framework (analytical, continuous, quantitative)**

This variant shifts learning from the organisational to the analytical level. It uses historical operating data, simulations and trend analyses to identify patterns and derive strategic adjustments from them. The focus is on continuous, data-based optimisation, which largely replaces manual reviews.

- **Core idea:** Time series analyses and modelling techniques identify long-term correlations between environmental conditions, sensor failures and system responses.
- **Technical implementation:** Use of statistical and AI-based analysis tools, clustering methods and simulation-based scenario evaluation.
- **Example:** Long-term data shows that sensors fail more frequently at certain humidity profiles. The AL then dynamically adjusts the ACL thresholds and optimises the maintenance frequency seasonally.
- **Reaction mode:** Adjustments are made continuously, automatically and data-driven with feedback to operational control levels.

This variant is highly complex and proactive, as it combines predictions and continuous improvement. Transparency is lower than with governance-oriented approaches, as decisions are based on statistical models. The degree of automation is high and the implementation effort considerable.

This variant is particularly suitable for systems with extensive data history and stable monitoring architectures. Therefore, this variant is particularly useful in complex, safety-critical CPSoS, for example in transport or energy systems, where adaptive strategies can be tested and optimised in simulation-based environments.

**Variant 5 – AI-supported knowledge integration and meta-learning (automated, networked, context-sensitive)**

This advanced variant implements the concept of a self-optimising, context-sensitive knowledge architecture. Here, the AL becomes an intelligent meta-learning platform that links experiences across systems and continuously adapts policies, thresholds and system strategies. It combines internal experiential knowledge with external sources to dynamically expand resilience-relevant knowledge.

- **Core idea:** AI models (e.g. LLMs and knowledge graphs) extract knowledge from heterogeneous data sources, recognise patterns and independently derive strategic adjustments.
- **Technical implementation:** Integration of semantic knowledge models, NLP (natural language processing) analyses and automated policy adjustment in real time.
- **Example:** Using data pools from several organisations, the system recognises that certain types of sensors systematically fail in warm, humid climates and automatically initiates an adjustment of the procurement policy and an update of operating guidelines.

- **Reaction mode:** Learning takes place decentral, almost in real time, with automatic feedback to operational control levels.

This variant achieves the highest level of complexity, proactivity and automation of all approaches, but is associated with reduced transparency and high implementation costs.

This variant is suitable for highly networked, data-rich system landscapes in which adaptive learning is understood as the strategic core of the resilience architecture. This variant unfolds its potential in highly dynamic, self-organising CPSoS – such as in autonomous production cells or intelligent network infrastructures – which learn resilient behaviour independently through continuous interaction.

The five variants of AL reflect different forms of organisational and data-driven learning ability. While reactive, manually structured approaches (variants 1 and 2) provide the framework for controlled experiential learning, particularly in safety-critical, highly regulated CPSoS, data-driven and AI-based mechanisms (variants 3 to 5) enable increasingly self-organised and context-adaptive knowledge integration. The variants are not arranged in a hierarchical line of development, but represent a spectrum of complementary learning strategies that can be combined in a targeted manner depending on system complexity, data availability and regulatory requirements.

Together with the ACL, the AL forms the strategic core of a closed resilience cycle in which short-term response and long-term system optimisation are interlinked.

The following section compares ACL and AL in order to evaluate their interaction in the context of resilience. In addition, the respective variants are discussed in terms of their contribution to achieving adaptive system capability, and potential fields of application and methodological limitations are identified.

## 6. Discussion

The proposed layers – Adaptive Coordination Layer (ACL) and Adaptation & Learning Layer (AL) – form complementary components within the extended resilience architecture. While the ACL addresses the operational level and enables short-term, data-driven adjustments during ongoing operations, the AL takes on the strategic function of knowledge consolidation and structural development. Both levels interact via continuous feedback: decisions made by the ACL provide input for the AL, whose results in turn refine the decision rules and thresholds of the ACL. This closes the gap between reactive recovery and proactive system adaptation that has existed in many frameworks to date.

In the following sections, the roles and interactions of both layers are first analysed with regard to the understanding of resilience, then the respective variants are compared with each other and possible application scenarios and limitations are discussed.

6.1 Comparison of ACL and AL

Although ACL and AL address different time horizons and functional logics, both pursue a common goal: the continuous maintenance and further development of system resilience. ACL acts as an operational control level that responds to identified risks in real time, prioritises countermeasures and dynamically allocates resources. It translates data-driven analyses, forecasts and status information directly into actions. In contrast, the AL functions as a strategic reflection level that learns from the results of the ACL, evaluates its effectiveness and supports the long-term adaptation of policies, thresholds and architectures. While the ACL ensures short-term stability and functional reliability, the AL creates the basis for sustainable learning and optimisation processes. This results in a structure based on division of labour but cyclically linked:

- Bottom-up, information and experience flow from the operational control levels into the AL.
- Top-down, the adjustments derived from this are fed back into the operational decision-making logic of the ACL.

This feedback loop combines short-term adaptation with long-term learning, thus closing a central gap in classic resilience architectures, in which reaction and learning are usually considered separately. The combination of ACL and AL allows resilience to be understood as a continuous, recursive process of mutual optimisation, in which the ACL stabilises operational system performance while the AL creates the structural conditions for future adaptability.

While the mutual interaction of ACL and AL thus forms the functional basis of a dynamic resilience architecture, the implementation variants that can be realised in both layers differ in terms of their level of adaptation, degree of automation and system transparency. The following section compares these variants in order to analyse the extent to which they address the resilience goals defined in the literature – in particular anticipation, resistance, recovery and adaptation – and in which system contexts an approximation to a holistically resilient system dynamic appears achievable.

6.2 Complementary Pathways to Resilience: Analysis of ACL and AL Variants

The implementation variants of ACL and AL described in the previous chapters illustrate different forms of adaptive control and organisational learning ability within CPSoS. Their contribution to system resilience can be evaluated using the target framework established in the literature, consisting of anticipation, resistance, recovery and adaptation. While earlier resilience models mostly focused on the static interpretation of individual dimensions, the variants presented here address the dynamic interaction of these capabilities across different time scales and levels of abstraction.

The early variants – such as rule-based ACL and manual post-incident learning of AL – focus on stability, recoverability and formalised traceability. They primarily secure the dimensions of resistance and recovery, as they respond to known events in a deterministic and verifiable manner. In highly regulated or safety-critical CPSoS, such as energy and transport systems, they thus form an essential basis for reliable system resilience, even if the ability to adapt independently remains deliberately limited here.

With the increasing integration of data-driven mechanisms – such as KPI-based feedback modules or continuous organisational learning loops – the contribution to resilience is expanded to include predictive and optimising components. These variants link resistance and recovery with elements of anticipation for the first time by detecting deviations at an early stage and deriving cyclically improved decision rules from them.

The advanced, AI-supported variants of both layers – in particular ML-based decision agents, data-driven adaptation frameworks and meta-learning approaches – ultimately shift the focus towards adaptation. They enable the system to independently generate new strategies based on continuous feedback and semantic knowledge integration. These characteristics thus approach the ideal of proactive, self-organising resilience that goes beyond mere reaction and establishes learning as an inherent system function.

However, this development should not be understood as a hierarchical improvement in quality, but rather as context adaptation: Variants with a higher degree of automation are not fundamentally more resilient, but require suitable framework conditions – such as reliable data quality, explainable decision-making logic and regulatory acceptance. The selection of a suitable combination of ACL and AL variants therefore depends on the complexity of the system, the availability of trustworthy data and the relationship between safety and security requirements and freedom of adaptation.

Overall, the comparison shows that resilience in CPSoS is not achieved through a single technical solution, but through the complementary interaction of different levels of adaptation: deterministic stability forms the basis, data-driven control expands responsiveness, and adaptive architectures anchor long-term adaptation. Only a balanced integration of these levels enables resilience to emerge as a dynamically stabilised, self-sustaining system capability.

6.3 Application Perspectives and System Contexts of Integrated Resilience Architecture Scenarios

While the possible applications of the individual variants have already been described in context, the following section focuses on the comprehensive integration of the ACL and AL concepts in different CPSoS domains. The aim is to identify typical system environments in which the interaction of both layers offers recognisable added value for resilience management. The practical relevance of the architecture is particularly evident when different domains have specific requirements in terms of response speed, traceability and learning ability.

The proposed ACL and AL concepts can be applied to different classes of CPSoS, whose requirements for resilience, transparency and adaptability vary considerably. Depending on the criticality, degree of regulation and data infrastructure, there are different areas of application for the respective variants of both layers.

In highly regulated, safety-critical infrastructures – such as energy supply, railway technology or aviation – traceability, auditability and stability are paramount. Deterministic and rule-based ACL variants in combination with organisationally anchored AL mechanisms that support structured post-incident analyses and governance-compliant learning processes are particularly suitable here. In these systems, resilience manifests itself primarily as controlled adaptability within clearly defined limits (bounded adaptivity).

Production and logistics systems with a medium degree of automation benefit from data-driven variants in which KPI-based feedback loops of the ACL are linked to continuous organisational learning processes of the AL. This allows operating parameters and maintenance strategies to be optimised iteratively without violating regulatory transparency requirements. In these scenarios, resilience acts as an operational optimisation capability based equally on process data and empirical knowledge.

In complex, distributed system landscapes – such as smart cities, intermodal transport systems or intelligent energy grids – advanced variants with a higher degree of automation are used. AI-based decision agents and data-driven adaptation frameworks enable context-dependent and partially autonomous control, while meta-learning mechanisms integrate cross-organisational knowledge. These systems are characterised by high dynamics and degrees of uncertainty, resulting in resilience as an emergent property from the interaction of many self-adaptive components.

Furthermore, the combination of ACL and AL offers potential for hybrid implementations in which different variants coexist in parallel. For example, a rule-based ACL can be used in an energy network to ensure network stability, while a data-driven AL takes over the strategic optimisation of maintenance intervals or network parameters. This hybrid coupling enables multi-scale resilience control that combines short-term responsiveness with long-term learning ability.

Overall, the application scenarios illustrate that the presented architecture is not limited to a specific system type, but offers a scalable framework concept. The selection of suitable variants depends largely on the balance between safety and security requirements, data availability and the desired degree of adaptation autonomy.

Although the application perspectives presented show that the combined architecture of ACL and AL can be used flexibly in different CPSoS domains, several key challenges arise from practical implementation. These challenges concern technical, methodological, and organisational aspects and critically determine the extent to which the theoretically formulated resilience objectives can be realised in real-world system environments.

6.4 Limitations and challenges

Although the application perspectives presented show that the combined architecture of ACL and AL has great potential in various CPSoS domains, there are several challenges associated with its practical implementation. These relate to technical, methodological and organisational aspects and define the framework conditions under which the theoretically formulated resilience objectives can be realised in real system environments.

Data quality and availability remain a key success factor. Resilience models can only be effective if the underlying input data is consistent, up-to-date and sufficiently representative of the specific domain. In heterogeneous system landscapes, this requires standardised interfaces, common data models and the use of robust sensor and monitoring infrastructures. However, advances in semantic data integration and edge computing show that this hurdle is becoming increasingly surmountable.

Another area of tension arises from the interpretability of data-driven decisions. With increasing automation, the complexity of models is growing, which makes traceability and regulatory acceptance more difficult. Approaches such as explainable AI, model-based transparency metrics or hybrid decision-making logic offer practical ways to ensure trust and auditability without limiting adaptability.

At the organisational level, the effectiveness of adaptive resilience architectures depends largely on the institutional anchoring of learning processes. The transition to data-driven, continuous learning requires not only technological integration, but also cultural openness to iterative adaptation and knowledge sharing. A gradual change is emerging in many industries – for example, through the establishment of data governance structures or interdisciplinary resilience teams.

Although implementation and validation costs represent a hurdle, they can be mitigated through modular architecture concepts and simulation-based evaluation. Initial pilot implementations of adaptive control and learning mechanisms can already be observed in industrial and energy domains. Early pilot projects such as DEIS [5], CAPRI [6] and GRIP [7] already point towards emerging forms of adaptive coordination, supporting the practical relevance of the proposed architecture.

7. **Conclusion & Future Work**

This paper has presented an extended resilience architecture for Cyber-Physical Systems of Systems (CPSoS) in which the newly introduced Adaptive Coordination Layer (ACL) and the redefined Adaptation & Learning Layer (AL) function as complementary layers for operationalising dynamic resilience. By formally distinguishing between the two layers and integrating them into existing architecture and governance models, the previous gap between reactive fault management and proactive system adaptation has been closed.

The central contribution lies in the conceptual linking of operational control logic (ACL) with strategic organisational and system learning (AL). This coupling makes it possible to understand resilience no longer as a static system property, but as a continuous, data-driven and self-reinforcing process. On a theoretical level, the model provides a framework for making resilience goals such as anticipation, resistance, recovery and adaptation measurable and controllable across all system levels. In practice, it opens up new opportunities to coordinate countermeasures in a context-sensitive manner, to institutionalise learning processes and thereby to strengthen both short-term responsiveness and long-term adaptability.

The proposed ACL can be understood as a scalable building block for future CPSoS resilience models. It acts as an operational interface between data-driven risk analysis, decision management and organisational learning, thus enabling adaptive control even in highly networked, heterogeneous system landscapes. In the future, it could be embedded in existing standard frameworks (e.g. ISO 31010 [15], NIST SP 800-160 Vol. 2 [10] or ENISA Guidelines [14, 28, 29]) in order to methodically anchor dynamic resilience assessment and continuous system adaptation.

Future research should focus on the empirical validation of the proposed architecture, in particular on the quantitative evaluation of the interaction between ACL and AL in real CPSoS environments. Of particular interest are metrics for measuring adaptive effectiveness, methods of explainable decision support (Explainable AI), and hybrid approaches that integrate human expertise into data-driven control logic. Furthermore, the simulation of failure and recovery behaviour under varying adaptation strategies appears to be a promising approach to quantifying and optimising the effectiveness of the proposed model.

This work thus answers the central research question of how resilience in CPSoS can not only be evaluated but also actively controlled. The concept presented here forms a theoretical and methodological basis for the next generation of resilient, adaptive CPSoS and emphasises that resilience must be understood not as a fixed state, but as an emergent property of continuous coordination and learning-based adaptation – a principle that is likely to significantly shape the development of future system architectures.

References

[1] E. Vogel, Z. Dyka, D. Klann, and P. Langendörfer, "Resilience in the Cyberworld: Definitions, Features and Models," *Future Internet*, vol. 13, no. 11, 2021, doi: 10.3390/fi13110293.

[2] E. Vogel and P. Langendörfer, "P10 -Enhancing Cyber-Resilience in Cyber-Physical Systems of Systems: A Methodical Approach," in *Poster*, Cottbus, May. 2024 - May. 2024, pp. 146–149.

[3] Z. Dyka, E. Vogel, I. Kabin, M. Aftowicz, D. Klann, and P. Langendorfer, "Resilience more than the Sum of Security and Dependability: Cognition is what makes the Difference," in *2019 8th Mediterranean Conference on Embedded Computing (MECO): Including ECYPS '2019 : proceedings-research monograph : Budva, Montenegro, June 10th-14th, 2019*, Budva, Montenegro, 2019, pp. 1–3.

[4] Z. Dyka, E. Vogel, I. Kabin, D. Klann, O. Shamilyan, and P. Langendörfer, "No Resilience without Security," in *2020 9th Mediterranean Conference on Embedded Computing (MECO)*, 2020, pp. 1–5.

[5] DEIS Consortium, *Dependability Engineering Innovation for Industrial Cyber-Physical Systems (DEIS Project Overview)*. European Union Horizon 2020 Programme, 2021. [Online]. Available: https://deis-project.eu/

[6] CAPRI Project Consortium, *Cognitive Automation Platform for Resilient Industry (CAPRI) – Project Summary and Pilot Results*. European Union Horizon 2020 Programme, 2022. [Online]. Available: https://www.cartif.es/en/capri-en/

[7] U.S. Department of Energy, *CPS Energy Grid Resilience and Innovation Partnership (GRIP) Pilot Project Report*. National Energy Technology Laboratory (NETL), 2024. [Online]. Available: https://netl.doe.gov/sites/default/files/2024-02/CPS%20Energy.pdf

[8] F. Björck, M. Henkel, J. Stirna, and J. Zdravkovic, "Cyber Resilience - Fundamentals for a Definition," in *New Contributions in Information Systems and Technologies*, 2015, pp. 311–316.

[9] V. Castano and I. Schagaev, *Resilient computer system design*. Cham: Springer, 2015. [Online]. Available: http://search.ebscohost.com/login.aspx?direct=true&scope=site&db=nlebk&AN=980100

[10] R. Ross, V. Pillitteri, R. Graubart, D. Bodeau, and R. McQuaid, "Developing Cyber-Resilient Systems: A Systems Security Engineering Approach," *National Institute of Standards and Technology*, 2021, doi: 10.6028/NIST.SP.800-160v2r1.

[11] Ron Ross, Richard Graubart, Deborah Bodeau, and Rosalie McQuaid, "Draft SP 800-160 Vol. 2, Systems Security Engineering: Cyber Resiliency Considerations for the Engineering of Trustworthy Secure Systems," 2021.

[12] E. Hollnagel, *Safety-II in practice: Developing the resilience potentials*. London, New York: Routledge, 2018.

[13] Bodeau, D., Graubart, R., Picciotto, J., McQuaid, R., "Cyber Resiliency Engineering Framework," *The MITRE Corporation, MITRE technical Report MTR110237*, 2011.

[14] European Union Agency for Cybersecurity (ENISA), "Best Practices for Cyber Crisis Management," 2024. [Online]. Available: https://www.enisa.europa.eu/publications/best-practices-for-cyber-crisis-management

[15] ISO 31010:2019, "Risk management: Risk assessment techniques," *International Organization for Standardization (ISO)*, 2019.

[16] Bodeau, D. Graubart, R., "Cyber Resiliency Assessment: Enabling Architectural Improvement," *The MITRE Corporation, MITRE technical Report MTR120407*, 2013.

[17] D. D. Woods, "Four concepts for resilience and the implications for the future of resilience engineering," *Reliability Engineering & System Safety*, vol. 141, pp. 5–9, 2015, doi: 10.1016/j.ress.2015.03.018.

[18] A. Kott *et al.*, "Approaches to Enhancing Cyber Resilience: Report of the North Atlantic Treaty Organization (NATO) Workshop IST-153," *ARL-SR-*. [Online]. Available: https://arxiv.org/pdf/1804.07651

[19] C. Berger, P. Eichhammer, H. P. Reiser, J. Domaschka, F. J. Hauck, and G. Habiger, "A Survey on Resilience in the IoT: Taxonomy, Classification and Discussion of Resilience Mechanisms," 2021. [Online]. Available: https://arxiv.org/pdf/2109.02328

[20] Ł. Jalowski, M. Zmuda, and M. Rawski, "A Survey on Moving Target Defense for Networks: A Practical View," *Electronics*, vol. 11, no. 18, p. 2886, 2022, doi: 10.3390/electronics11182886.

[21] M. Salayma, "Risk and threat mitigation techniques in internet of things (IoT) environments: a survey," *Front. Internet Things*, vol. 2, 2024, doi: 10.3389/friot.2023.1306018.

[22] S. Rauti and S. Laato, "Enhancing resilience in IoT cybersecurity: the roles of obfuscation and diversification techniques for improving the multilayered cybersecurity of IoT systems," *Data & Policy*, vol. 6, 2024, doi: 10.1017/dap.2024.84.

[23] C. Kilinc *et al.*, "JADE: Data-Driven Automated Jammer Detection Framework for Operational Mobile Networks," in *IEEE INFOCOM 2022 - IEEE Conference on Computer Communications*, London, United Kingdom, 2022, pp. 1139–1148.

[24] Y. Ye, Z. Nie, X. Liu, F. Xie, Z. Li, and P. Li, "ROS2 Real-time Performance Optimization and Evaluation," *Chin. J. Mech. Eng.*, vol. 36, no. 1, 2023, doi: 10.1186/s10033-023-00976-5.

[25] V. Stefanova-Stoyanova and P. Danov, "Comparative Analysis of Specialized Standards and Methods on Increasing the Effectiveness and Role of PDCA for Risk Control in Management Systems," in *2022 10th International Scientific Conference on Computer Science (COMSCI)*, 2022, pp. 1–4.
[26] L. L. Klein, S. I. D. de Pádua, R. Gera, K. M. Vieira, and E. C. H. Dorion, "Business process management effectiveness and maturity through lean management practices: the Brazilian federal police experience," *IJLSS*, vol. 14, no. 2, pp. 368–396, 2023, doi: 10.1108/IJLSS-07-2021-0125.
[27] J. C.M. Franken, D. H. van Dun, and C. P.M. Wilderom, "Kaizen event process quality: towards a phase-based understanding of high-quality group problem-solving," *International Journal of Operations & Production Management*, vol. 41, no. 6, pp. 962–990, 2021, doi: 10.1108/IJOPM-09-2020-0666.
[28] European Union Agency for Cybersecurity (ENISA), "NIS2 Technical Implementation Guidance: ENISA Report," 2025. [Online]. Available: https://www.enisa.europa.eu/publications/nis2-technical-implementation-guidance?utm_source=chatgpt.com
[29] R Andrew Paskauskas, "ENISA: 5G design and architecture of global mobile networks; threats, risks, vulnerabilities; cybersecurity considerations," *Open Research Europe*, vol. 2, p. 125, 2023, doi: 10.12688/openreseurope.15219.3.